\documentclass[aps,prl,twocolumn,amsmath,amssymb]{revtex4}
\usepackage{amsfonts}
\usepackage{graphicx}
\usepackage{amsbsy}
\usepackage{hyperref}
\usepackage{dcolumn}
\usepackage{bm}

\begin{document}
\draft \noindent \textbf{Comment on ``Topological Oscillations of
the Magnetoconductance in Disordered GaAs Layers''}
\vspace{0.5cm}

In a recent Letter, Murzin {\em et. al.}~\cite{Murzin}
investigated ``instanton effects" in the magneto resistance data
taken from samples with heavily Si-doped GaAs layers at low
temperatures ($T$). This topological issue originally arose in the
development of a microscopic theory of quantum Hall effect some 20
years ago.~\cite{Pruisken} The investigations by Murzin {\em et.
al.}, however, do not convey the correct ideas on scaling that
have emerged over the years in the general theory of quantum
transport. The physical idea that has been overlooked by Murzin
{\em et. al.} is that the {\em macroscopic} conductances measured
at finite $T$ can in general be very different from {\em ensemble
averaged} conductances at $T=0$ that one normally uses for
theoretical purposes. The concept of {\em conductance
fluctuations} implies, for example, that a complete knowledge of
the macroscopic conductances at finite $T$ necessarily involves a
detailed knowledge of the complete conductance {\em distributions}
at $T=0$.~\cite{CohenPruisken} This is one of the reasons why the
explicit results of phenomenological approaches~\cite{Dolan}
quoted in Ref.~\cite{Murzin} generally lack any kind of
microscopic justification.

Unlike the expressions derived in Ref.~\cite{Murzin}, the general
scaling behavior of macroscopic quantities is obtained by using
well known methods of quantum field theory (method of {\em
characteristics}).~\cite{ZinnJustin,PruiskenWang} Denoting the
{\em ensemble averaged} longitudinal and Hall conductances by
$\sigma_0$ and $\sigma_H$ respectively then the {\em measured}
quantities $G_{ij}= G_{xx}, G_{xy}$ at finite $T$~\cite{Murzin}
obey the differential
equations~\cite{BaranovBurmistrovPruisken,PruiskenBurmistrov}
\begin{equation}
 \left[ \beta_\sigma \frac{\partial}{\partial \sigma_0} + \beta_H
 \frac{\partial}{\partial \sigma_H} + (2 + \gamma)zT
 \frac{\partial}{
 \partial zT} \right] G_{ij} = 0 .
\end{equation}
Here, the $\beta_\sigma$, $\beta_H$ and $\gamma$ are the
renormalization group functions associated with $\sigma_0$,
$\sigma_H$ and the singlet interaction amplitude $z$ respectively.
Explicit results in the weak coupling regime have been obtained
only recently~\cite{PruiskenBurmistrov}
\begin{eqnarray}
 \beta_\sigma &=& \hspace{0.35cm}-\frac{2}{\pi} -\frac{b_1}{\sigma_0} - D_0 \sigma_0^2
 e^{-2\pi\sigma_0} \cos 2\pi \sigma_H \\
 \beta_H &=& \hspace{1.9cm} - D_0 \sigma_0^2
 e^{-2\pi\sigma_0} \sin 2\pi \sigma_H \\
 \gamma &=& -\frac{1}{\pi\sigma_0} -
 \frac{c_2}{\sigma_0^2} - \frac{D_0}{6} \sigma_0 e^{-2\pi\sigma_0} \cos 2\pi
 \sigma_H
\end{eqnarray}
where $b_1 \approx 0.66$, $c_2 \approx 0.47$ and $D_0 \approx
13.56$ are numerical constants. In the regime $e^{-2\pi\sigma_0}
\ll 1$ we can write~\cite{Murzin}
\begin{equation}
 G_{xx}\approx g_0 - g_1 \cos 2\pi \sigma_H,\,\,
 G_{xy}\approx \sigma_H - h_1 \sin 2\pi \sigma_H  \label{a1}
\end{equation}
where the functions $g_0$, $g_1$ and $h_1$ depend on $T$ only
through the scaling variable $X$
\begin{equation}
g_0 = g_0 (X) ,\quad X = z T \xi^2 M_0 ,\quad \xi = l_0
\sigma_0^{-\pi^2 b_1/4} e^{\pi\sigma_0/2}. \label{g0}
\end{equation}
Here, $\xi \gg l_0$ is the correlation length, $l_0$ a microscopic
length and $M_0 = \sigma_0^{-1/2}[1+ \pi (2 c_2-b_1)/(2\sigma_0) +
\mathcal{O} (\sigma_0^{-2})]$.~\cite{ZinnJustin} On the other
hand, we find
\begin{equation}
g_1 = [f_\sigma(g_0) - f_\infty(\sigma_0)] \pi X g_0^\prime,\,\,\,
h_1 = f_H (g_0) - f_\infty (\sigma_0) \label{a3}
\end{equation}
with $f_\infty (\sigma_0) = [(D_0/4) \sigma_0^2 + \mathcal{O}
(\sigma_0)] e^{-2\pi\sigma_0}$. In the limit $G_{xx}$, $g_0
\rightarrow \infty$ one expects that $g_0 (X) \to (1/\pi) \ln
X$~\cite{BaranovBurmistrovPruisken} but no first principle
computation of $f_{\sigma , H} (g_0)$ exists.
Experimentally~\cite{Murzin} $G_{xx} \approx g_0 =\mathcal{O} (1)$
such that both the convergence of the series in
$\sigma_0^{-1}$~\cite{BaranovBurmistrovPruisken} and the
occurrence of broad conductance distributions~\cite{CohenPruisken}
now complicate the problem. These complications are clearly
reflected by the fact that the $g_0 (X)$ data in the range of
experimental $T$~\cite{Murzin} do not truly display the
aforementioned asymptotic behavior in $X$. In summary,
\begin{equation}
 G_{xy}(T) \approx \bar{\sigma}_H - \left[ f_H (g_0) - f_H (\bar{g}_0)
 \right]
 \sin 2\pi \bar{\sigma}_H  \label{a4}
\end{equation}
with $\bar{\sigma}_H$, $\bar{g}_0$ denoting the values of
$G_{xy}$, $g_0$ at a fixed $T$, say $T_0$, is a very general
result that is predicted by the theory~\cite{PruiskenBurmistrov}.
The most important feature of the experimental result $f_H (g_0)
\approx 7.6 e^{-2\pi g_0}$ for $g_0 = 1.4 -- 1.9$~\cite{Murzin} is
the exponential dependence on $g_0$ which actually has a much more
general significance in quantum field theory over a much larger
range in $g_0$.~\cite{Pruisken,PruiskenBurmistrov}

\vspace{0.2cm}
\noindent
A.\,M.\,M.\,Pruisken$^1$ and
I.\,S.\,Burmistrov$^{1,2}$\\
{\small ${}^1$Institute for Theoretical Physics, University
of Amsterdam, \\
\hphantom{${}^1$}Valckenierstraat 65, 1018 XE Amsterdam, The
Netherlands\\
${}^2$L.\,D.\,Landau Institute for
Theoretical Physics,\\
\hphantom{${}^2$}Kosygina street 2, 117940 Moscow, Russia }

\end{document}